\documentstyle[preprint,12pt, epsfig]{aastex}
\begin{document}
\setcounter{page}{1}

\title{Unusual Properties of X-Ray Emission near the Galactic Center}


\author{Yasuo {\sc Tanaka} \\
{\it The Institute of Space \& Astronautical Science, 3-1-1 Yoshinodai, 
Sagamihara, Kanagawa 229-8510,}\\
{\it Max-Planck Institut f\"{u}r Extraterrestrische Physik, D-85748 
Garching, Germany}\\
{\it E-mail(Y.T.): ytanaka@mpe.mpg.de}
\\[6pt]
Katsuji {\sc Koyama}\\
{\it Dept. of Physics, Kyoto University, Sakyo-ku, Kyoto 606-01}\\  
Yoshitomo {\sc Maeda}\\
{\it Dept. of Astronomy \& Astrophysics, Penn State University, 
University Park, PA 16802-6305, U.S.A.}\\
and\\
Takashi {\sc Sonobe}\\
{\it  Fujitsu Ltd., 4-1-1 Kamikodanaka, Nakahara-ku, Kawasaki, Kanagawa
211-8588}
}


\slugcomment{\protect\small To be appeared as Publ. Astron. Soc. Japan 52, L25-L30 (2000)}

\shorttitle{X-ray Emission near the Galactic Center}

\shortauthors{Y. Tanaka et al.}

\begin{abstract}

The X-ray spectrum in a $1^{\circ}\times1^{\circ}$ region of the Galactic 
center observed with the ASCA satellite is examined in detail, following 
the first report by Koyama et al. (1996, AAA 65.155.208). 
The observed spectrum contains prominent emission lines from 
helium-like and hydrogen-like ions of various elements, and is essentially 
the same all over the region. 
If the observed spectrum is thermal emission from hot plasmas, 
it requires multi-temperature plasma components, each at a different degree of 
ionization and with a different amount of absorption. 
The absence of adiabatic cooling and of systematic changes in the 
degree of ionization over the region is against the Galactic center origin 
of hot plasmas. 
A significant broadening of the helium-like and hydrogen-like 
iron K-lines is confirmed. The line width corresponds to a rms velocity of 
$\sim3300$ km~s$^{-1}$, which far exceeds the sound velocity in a plasma of 
$kT\sim14$ keV measured with the Ginga satellite. 
These facts cast doubt on a thermal origin of the observed X-ray emission.

\end{abstract}


\maketitle
\thispagestyle{headings}

\section{Introduction}
The Galactic center (GC) region is unique in various aspects (for reviews 
see Genzel et al. 1994; Mezger et al. 1996).  
Eckart and Genzel (1996) discovered the presence of a massive black hole 
of $\sim2.5\times10^6$ $M_{\odot}$ coincident with the compact radio source 
Sgr A$^*$.  
Although the central black hole is currently inactive, the complex features 
in the GC region may reflect its past activities.

The first detailed X-ray imaging and spectroscopic observations of the 
GC region were carried out with the ASCA satellite. The first results were 
published by Koyama et al. (1996) (hereafter referred to as Paper I). 
As shown in Paper I, the X-ray brightness structures are strikingly similar  
to the radio brightness structures (see figure 1b of Paper I), strongly 
suggesting that the X-ray emission and the radio emission are physically 
connected in one way or another.

The ASCA observation also provided an X-ray spectrum of the 
GC region with unprecedented resolution (figure 2 in Paper I). 
The general properties of the X-ray spectrum were discussed in Paper I, 
based on models of thermal emission from hot plasmas. 
It was also mentioned in Paper I that the helium-like and hydrogen-like 
iron K-lines apparently show Doppler broadening.  
Since this is crucial for understanding the origin of the 
observed X-ray emission, a confirmation and further study of this line 
broadening is very important.   
In this paper, we re-examine the spectral data from the ASCA observation 
of the GC region, and discuss the results and implications.  

\section{Observed Spectrum}
ASCA carries two types of imaging spectrometers, GIS and SIS 
(see Tanaka et al. 1995). In this work, we deal only with the SIS 
data for the higher spectral resolution. The SIS consists of two detectors, 
S0 and S1, respectively placed at the foci of two co-aligned X-ray 
telescopes. Both comprise four CCD chips (C0--C3) and cover a common 
square field of view of $\sim22'\times22'$. 
The ASCA observation of the GC region covered approximately 
$1^{\circ}\times1^{\circ}$ ($\sim150\times150$ pc$^2$) of the GC in 
a mosaic of eight SIS fields with the pointing positions listed
in table 1 (see figure 1a of Paper I). 
At the time of this observation (1993 September 30--October 5), 
the measured FWHM was approximately 140 eV for the iron K${\alpha}$-line 
at 6.4 keV (or $\sigma\approx60$ eV).
Data were screened according to the normal selection criteria. 
In addition, we required the cut-off rigidity to be greater than 8 GV. 
The exposure time for useful data of each field was between 14 ks and 
20 ks, depending on the pointing position. 

\begin{table}
\begin{center}
Table 1. The Galactic coordinates of the eight pointing positions  

\begin{tabular}{cll|cll}
\hline\hline
Position& $l^{\rm II}$ &$b^{\rm II}$ &Position &$l^{\rm II}$ &$b^{\rm II}$\\ 
\hline
1 &359.97 &+0.01 &5 &359.83& $-$0.50 \\
2 &0.33   &+0.22 &6 &359.73& $-$0.15\\
3 &0.51 & $-$0.08 &7 &359.47&+0.09\\
4 &0.15 & $-$0.28 &8 &359.81&+0.31\\
\hline
\end{tabular}
\end{center}
\end{table}

\begin{figure}[h]
\begin{center}
\epsfig{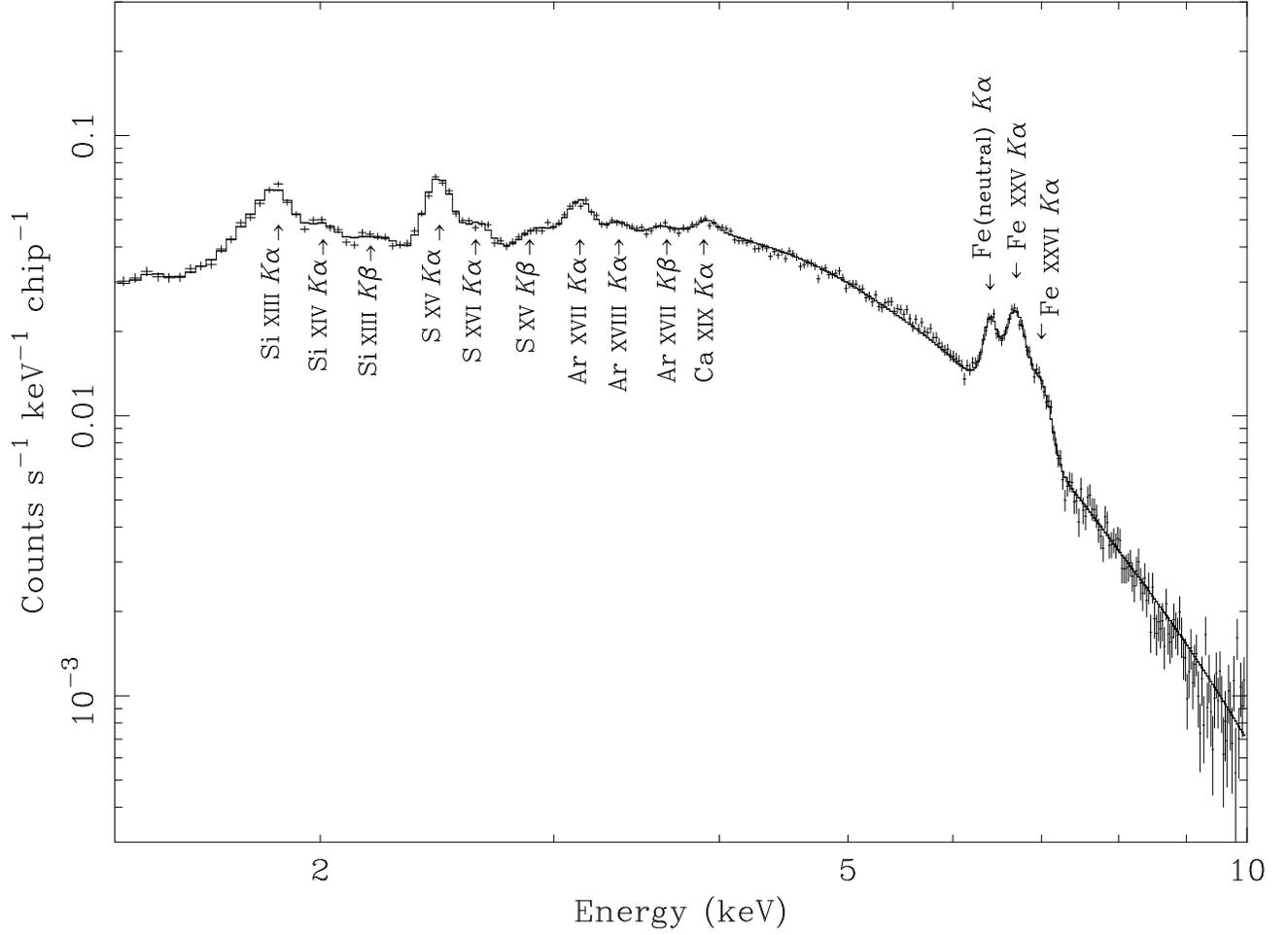}
\caption[]{Spectrum for the entire GC region observed. The discrete source 
contributions have been removed. The identified emission lines are indicated, 
respectively. The best-fit model is also included.  

\label{figure:1}
}
\end{center}
\end{figure} \normalsize

\begin{figure}[t]
\begin{center}
\epsfig{file=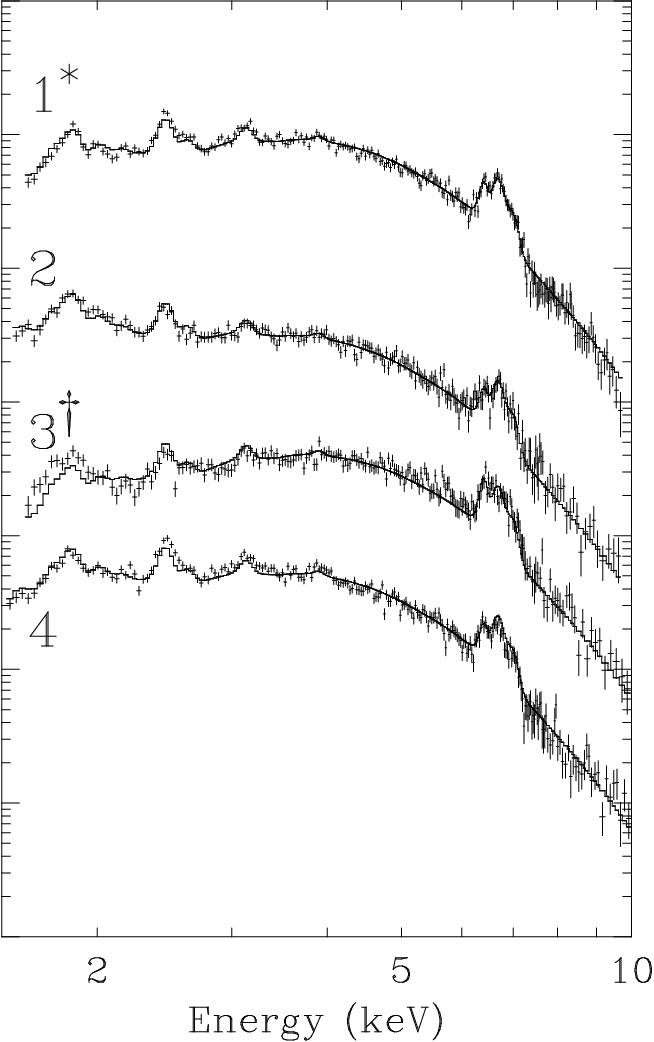,width=2.65in,angle=-90}
\epsfig{file=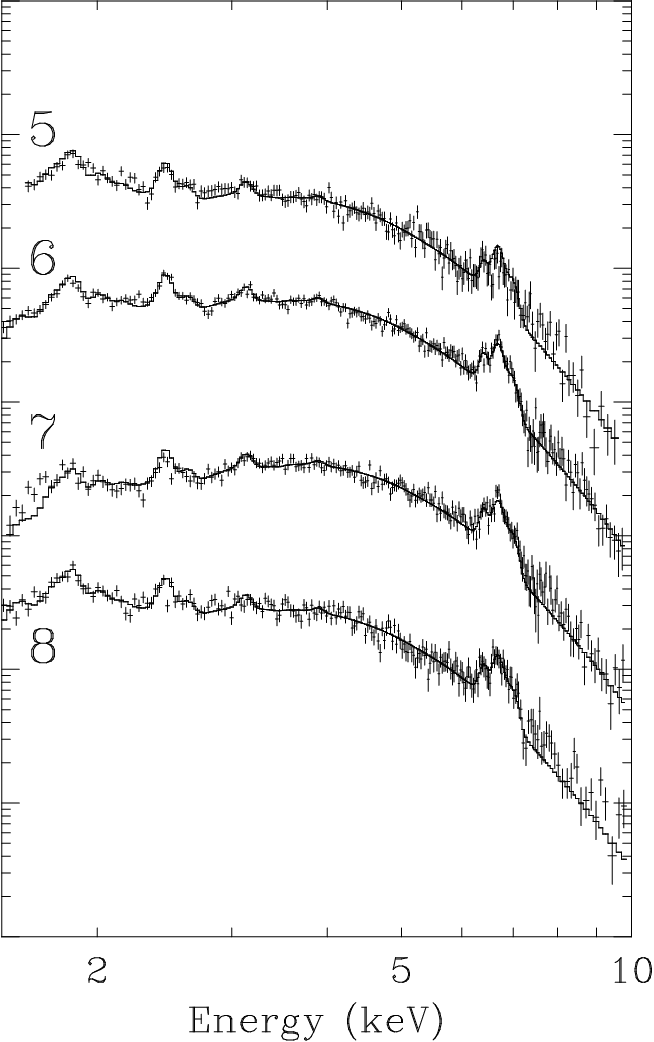,width=2.65in,angle=-90}
\caption[]{Observed spectra at eight pointing positions. The unit of the 
vertical axis is arbitrary. \\ 
(*): The spectrum for Position 1 excludes the Sgr A region.\\
($\dagger$): The spectrum for Position 3 excludes an area of Sgr B2.

\label{figure:2}
}
\end{center}
\end{figure} \normalsize

We constructed spectrum for each of the eight fields, after removing 
the detected point sources. The background 
data were obtained from the work by Gendreau et al. (1995) based on  
observations of high galactic latitude regions. The background consists 
of the cosmic X-ray background (CXRB) and the instrument background. 
The latter contains fluorescence lines of iron and nickel, of which the 
nickel line is the strongest. 
The background was determined by adjusting the instrument background 
level so that the nickel line (7.48 keV) intensity would be equal to that 
in the observed spectrum. To do this, we summed all eight spectra 
together to allow an accurate measurement of the nickel line intensity. 
We have not applied galactic absorption to the CXRB spectrum. Though the 
low-energy absorption toward the GC region is quite significant, the effect 
is small above 2 keV owing to a good S/N ratio, except for the range above 
7 keV. Thus, the derived spectrum for the entire GC region is shown in 
figure~1 (the same as figure 2 in Paper I). 

The observed emission lines are identified with K${\alpha}$-lines of 
helium-like and hydrogen-like ions of silicon, sulphur, argon, calcium, and 
iron. The helium-like K${\beta}$-lines of silicon, sulphur, and argon are 
also significantly visible, as marked respectively in figure~1. 
In addition, a significant 6.4-keV fluorescent iron K${\alpha}$-line is 
present. 
The best-fit model comprising emission lines and a continuum is also shown 
in Fig. 1. The continuum is expressed by a power-law with a 
photon index of $\sim1.2$. If approximated by a thermal bremsstrahlung 
spectrum, the electron temperature is higher than 10 keV, but unbound 
because of the limited energy band. Previously, Yamauchi et al. (1990) 
obtained a temperature of $14\pm1$ keV for the same GC region from  
observations over a wider energy band up to 20 keV with the Ginga satellite. 
We use this temperature for thermal emission in later discussions. 

For the purpose of supplementing Paper I, the spectrum for each individual 
field is presented in figure~2, in which the vertical scale is successively 
shifted. 
The background determined for the summed spectrum was subtracted from each 
observed spectrum. As noted in Paper I, these spectra are very similar 
to each other, except for the region of Srg B2, which shows a distinctly  
different spectrum, and is not included in figure~2. 
This region exhibits a signature of strong X-ray reflection with a very 
hard continuum and a pronounced 6.4 keV line (see Murakami et al. 2000). 
The model spectrum drawn on each spectrum is the best-fit to the summed 
spectrum (shown in figure~1), except that the absorption, normalization and 
the 6.4-keV line intensity are individually determined. 
It is noticed that the fit is not good near the silicon line ($\sim1.9$ keV) 
for Position 3 and 7, where the absorption is particularly strong. 
The fit improves by considering a distributed absorber within the emission 
region instead of placing all of the absorber in front of the emission region. 
For a further test, the equivalent widths of the prominent lines 
(helium-like Si, S, Ar, Fe, and hydrogen-like Fe) were individually determined 
for each field. 
The measured equivalent widths for each element were found to be the same 
within $\pm30$\% over the entire region observed, except for a systematic 
trend that the iron line intensity slightly decreases with the Galactic 
latitude.    
Thus, one can conclude that the intrinsic spectrum is essentially identical 
over the entire $1^{\circ}\times1^{\circ}$ area around the GC. 
This also justifies the summation of all individual spectra to obtain the best 
statistical accuracy (figure~1).

\section{Iron Line}

\begin{table}[ht]
\begin{center}
Table 2. Energies of the iron K-lines and the line widths
\medskip

\begin{tabular}{lcl}
\hline\hline
\noalign{\smallskip}
true (keV) & observed (keV) & line width $\sigma$ (eV) \\ 
\hline
\noalign{\smallskip}
\multicolumn{3}{l}{Step 1 ---------} \\
\ 6.40 & 6.42 (6.41--6.43) & 30 (0--47) \\
\ 6.70 & 6.72 (6.70--6.73) & 93 (69--120)\\
\ 6.97 & 7.00 (6.96--7.04) & 88 (62--120)\\
\hline

\multicolumn{3}{l}{Step 2 ---------} \\
\ 6.40 & 6.42 (6.41--6.43) & 27 (0--44)\\
\ 6.70 & 6.72 (6.71--6.73) & 78 (70--85)\\
\ 6.97 & fixed at $1.04\times$above  & = above\\
\hline
\ 7.48 * & 7.53 (7.51--7.56) & 36 (0--69) * \\ \hline
\hline

\end{tabular}
\end{center}
\smallskip
{\small  
Step 1: Results obtained from all CCD chips. \\
Step 2: Results from the data, excluding S0C2 and S1C0 and Sgr A and 
the Position 3 field, where the two line energies are linked (see text).\\}
$*$ {\small 
The fluorescent K-line of nickel in the instrumental background. 
The line width error is large due to poor statistics.\\
The numbers in parentheses are the 90\% confidence limits.}

\end{table}

\begin{figure}[ht]
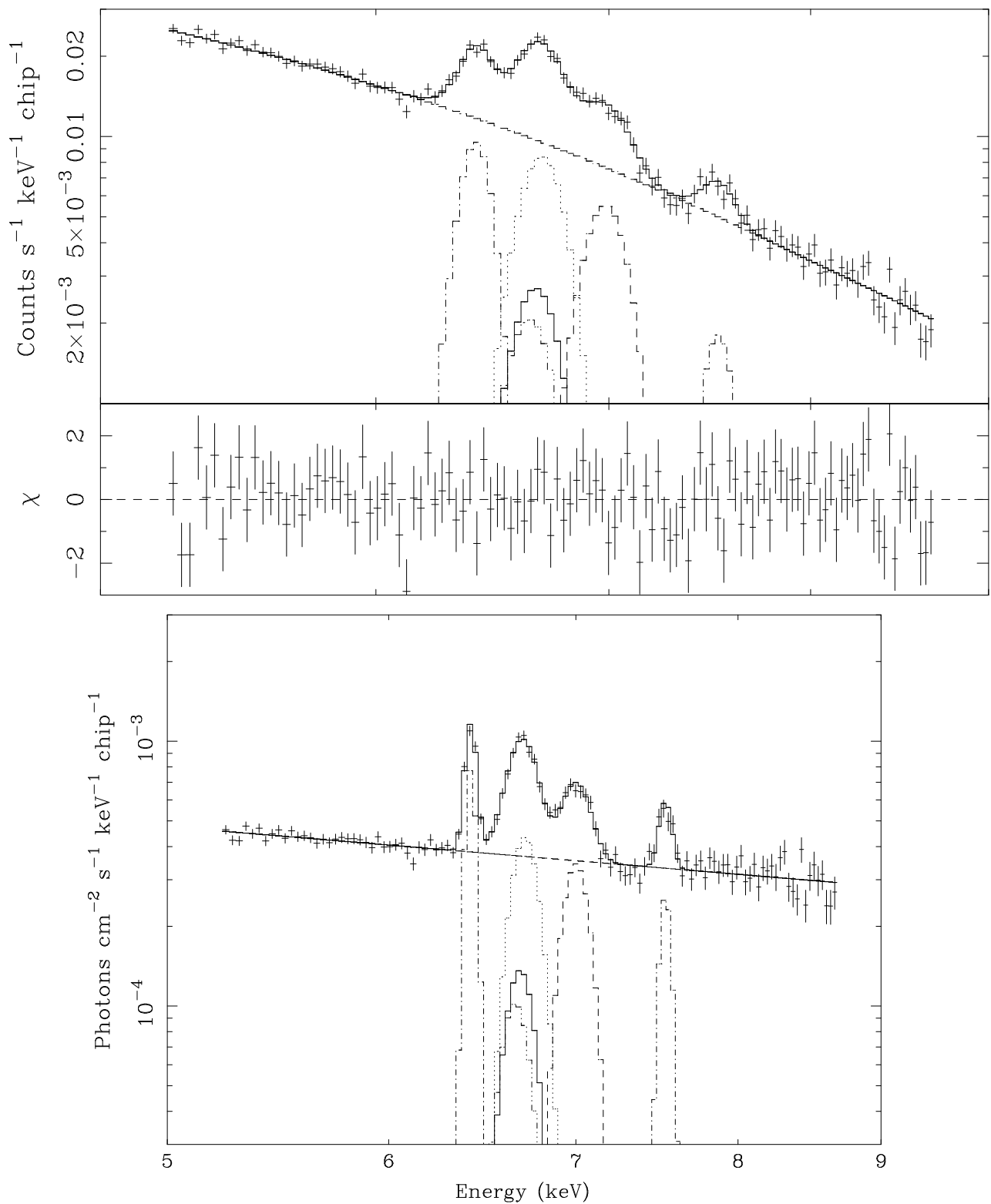

\begin{center}
\epsfig{file=./fig3_upper.ps,width=4in,angle=-90}\\
\vspace*{8pt}
\epsfig{file=./fig3_lower.ps,width=4in,angle=-90}\\
\caption[]{ Upper panel: Best-fit to the observed iron line complex obtained 
with six qualified chips (see text), excluding the Sgr A region and the 
Position 3 field. The background is not subtracted, and the instrumental 
nickel line (7.48 keV) is also present. 
Lower panel: Same as the upper panel, but deconvolved with the energy response 
function of the SIS to approximately depict the true line profiles. 
\label{figure:3}
}
\end{center}
\end{figure} \normalsize

The iron emission line complex consists of at least three lines, a 
fluorescent K${\alpha}$-line from low-ionization iron atoms (below XVII stage),  
K${\alpha}$-lines from helium-like and hydrogen-like ions. A model fitting 
is performed to the observed eight-field summed spectrum (before background 
subtraction to preserve the best statistics) in the range 5--9 keV. 
We also took into account that the helium-like K${\alpha}$-line is a triplet 
(6.64 keV, 6.67 keV, and 6.70 keV), which are not resolved with the SIS energy 
resolution. 
For possible offsets in the energy scale, we kept 
the observed energy of the 6.70-keV line as a free parameter, while 
the energy ratios of the other two lines to 6.70 keV were fixed. 
The intensity ratios of the three lines are also fixed, taken from the 
calculation for $kT=10$ keV by Masai (1984). 
The hydrogen-like line at 6.97 keV is a singlet, and the line 
energy was also kept free for fitting. A satisfactory fit was obtained 
with $\chi^2=92$ for 106 d.o.f. The thus-determined line energies and the line 
widths (Gaussian $\sigma$) 
are given in table 2 ``Step 1''. 
The measured line energies are in excellent agreement 
with the true values within a very small calibration uncertainty.
Including the weak fluorescence K${\beta}$ line (7.05 keV) did not 
change the result. 
If, as a reasonable assumption, we fixed the ratio of the energies of the 
hydrogen-like and helium-like lines at that of the true values 
(6.97 keV/6.70 keV), keeping the observed 6.7-keV line energy free, 
and requiring the same width for both lines, we obtained a line width of 92 eV 
with a 90\% confidence range of 85--101 eV. 

The 6.4-keV line is most probably the fluorescence line emitted from cool 
matter, and is hence an intrinsically narrow line. 
(Note that the contribution from 
the 6.4-keV line in the internal background is negligibly small.) 
The observed line width of 30 eV is considered to be due to systematic offsets 
of the energy scale among eight SIS chips, since the 6.4-keV iron line and 
the 7.5-keV nickel line in the internal background also show a $\sim30$ eV 
width.


As a next step for scrutinizing any possible systematic effects, we examined 
the data from each of the eight chips individually with respect to the 
energy calibration and resolution. 
To perform the examination, we excluded the Sgr A region, which is 
exceptionally bright as well as heavily absorbed, and also Position 3, 
which includes the giant molecular cloud Sgr B2, for the previously explained 
reason. All of the spectra in other fields were 
added for each SIS chip. Then, each summed spectrum was fitted individually 
in the same way as described above. 
As a result, we have found that two chips, S0C2 and S1C0, exhibit a weaker 
(hence less-defined) 6.4-keV line and somewhat broader helium- and 
hydrogen-like lines than the other six chips. 
For the rest, the energy scale is confirmed 
to be accurate within $\pm20$ eV. The data from these two chips are 
excluded, and all other data are summed again. 

The same model as before was used for fitting. The result of the fit  
($\chi^2=110$ for 108 d.o.f) is shown in figure~3, and the best-fit line 
energies and widths are given in table 2 ``Step 2''.
While the line-energy values remain the same as before, the widths of 
the 6.4-keV line and the helium-like and hydrogen-like lines are slightly 
reduced to 27 eV and 78 eV, respectively. 
The difference between the two widths is statistically quite significant. 
We thus conclude that the broadening of the helium-like and hydrogen-like 
lines is real. 
Considering the width of the 6.4-keV line to be systematic, we obtained a 
Gaussian width, $\sigma$, of $73\pm14$ eV (90\% confidence error) for these 
lines.
  
We investigated whether or not this broadening could result from a spatial 
variation of the line energy. The line energies were determined for each 
individual field, i.e. for the spectra shown in figure~2, except for Postion 3. 
As a result, the energies of the 6.7-keV line are all consistent with being 
constant with a rms dispersion of $\sim11$ eV. 
Therefore, for a possible spatial energy variation, the scale size of the 
variation should be smaller than an SIS field of $20'\times20'$.

\section{Discussion}
The observed properties of the X-ray emission near to the GC are unusual in 
several respects. 
The spectrum is very hard. If it is thermal emission, 
the temperature could be $\sim14$ keV according to the previous Ginga 
observation (Yamauchi et al. 1990), which is far higher than that of 
young supernova remnants. 
The presence of strong lines from silicon, sulphur and argon at this 
temperature requires either non-equilibrium ionization ({\it nei}) or 
multi-temperature components, or both. Employing the {\it nei} plasma model  
developed by Masai (1984), we find that a single-temperature {\it nei} 
plasma cannot account for the observed spectrum. 

In the observed spectrum, not only the helium-like and hydrogen-like 
K${\alpha}$-lines of silicon, sulphur, argon, and iron, but 
also the helium-like K${\beta}$-lines of silicon, sulphur, and argon are  
resolved. The line intensity ratio, $I$(H-K${\alpha}$)/$I$(He-K${\alpha}$), 
of each element is a function of the electron temperature, $T_{\rm e}$, and 
the degree of ionization determined by $n_{\rm e} t$, where $n_{\rm e}$ is 
the electron density and $t$ the age of the plasma. Here, He- and H- denote 
helium-like and hydrogen-like, respectively.
However, the line-intensity ratio $I$(He-K${\beta}$)/$I$(He-K${\alpha}$) 
is a sole function of $T_{\rm e}$. Therefore, for silicon, sulphur, and argon, 
the measured line intensities, $I$(He-K${\alpha}$), $I$(H-K${\alpha}$) and 
$I$(He-K${\beta}$), constrain both $T_{\rm e}$ and $n_{\rm e} t$. 

We attempted to see this constraint qualitatively for the lines of silicon, 
sulphur and argon separately, using the {\it nei} plasma code by Masai (1984).
For each element, the energy range was limited to cover only those lines from 
the element. We assumed that the lines of an element come from a 
single-temperature plasma. The results are listed in table 3. 
Although the errors are relatively large, because of coupling 
between the parameters, the results indicate that each element 
requires a different temperature than others, hence demonstrating the 
necessity of multi-temperature components. 

In reality, the situation is more complicated, since various temperature 
components contribute to the lines of each element. 
We attempted to reproduce the observed spectrum with a multi-component 
{\it nei} plasma model, but failed to obtain a reasonable fit. 
Besides, a fundamental problem exists. In order to reproduce the observed 
hard continuum, each temperature component inevitably requires a different 
amount of absorption than others, which is artificial and quite unrealistic. 
This is a serious problem in a multi-temperature plasma 
interpretation.

\begin{table}
\begin{center}
Table 3. Plasma temperature, $n_{\rm e} t$ and abundance for 
the individual elements
\medskip

\begin{tabular}{lllc}
\hline\hline
\noalign{\smallskip}
Element &  $kT_{\rm e}$ (keV) & \ \ log $n_{\rm e} t$  & Abundance (solar)\\ 
\hline
\noalign{\smallskip}
Si &  $1.5\pm0.3$ & $11.1\pm0.2$ & $0.33\pm0.4$ \\
S  &  $2.2\pm0.1$ & $11.1\pm0.1$ & $0.29\pm0.4$ \\
Ar &  $6.2\pm0.7$ & $11.1\pm0.3$ & $0.23\pm0.4$ \\
Fe & 14 (fixed)   & $11.8\pm0.1$ & $0.42\pm0.3$ \\
\hline

\end{tabular}
\end{center}

* {\small The errors are the 90\% confidence limits.}

\end{table}

We have confirmed that the helium-like and hydrogen-like iron K-lines are 
broadened with $\sigma\approx73$ eV. This corresponds to a rms velocity 
dispersion of $\sim3300$ km s$^{-1}$ of iron ions. Broadening of the lines 
of lighter elements is not conclusive, because of decreasing resolution at 
lower energies. As discussed in Paper I, it is quite unlikely that the 
observed velocity dispersion is due to the thermal motion of ions, since it 
corresponds to a temperature of $\sim3$ MeV. It is most probably 
due to bulk motions.  

As shown in the previous section, no spatial variation is found in the iron 
line energy within a scale of at least $20'$ ($\sim50$ pc), implying that bulk 
motions at various directions are taking place within this scale.  
This is against large-scale systematic motion, such as rotation 
around the GC or directed (e.g. radial) streaming. If we assume the motion 
to be nearly random, the actual velocity dispersion would be even larger, 
$\sim5000$ km s$^{-1}$, since the line width represents the line-of-sight 
velocity component. 

If the observed X-rays are thermal emission, the plasma pressure is 
$\sim7\times10^{-9}$ erg cm$^{-3}$ for a temperature of $\sim14$ keV and a 
density of $\sim0.3$ electron~cm$^{-3}$, as derived from the emission measure 
(see Paper I).  
If magnetic fields on the order of mGauss are present, these plasmas are 
contained by the magnetic fields and will propagate along the field lines. 
This picture is consistent with the observed similarity 
in the brightness structure between X-rays and radio, the latter being 
synchrotron radiation by magnetically-trapped high-energy electrons. 
On the other hand, if the hot plasma were produced at the GC, as considered 
in Paper I, the plasma would be adiabatically cooled as it expands. 
This cooling is expected to be quite significant, but it is not observed. 
The observed spectrum appears to be the same everywhere. A possibility was 
discussed in Paper I to maintain the electron temperature through energy 
transfer from ions that carry most of the energy in an {\it nei} plasma. 
However, the collisional ionization would proceed during propagation. 
If the plasma moves at the inferred speed to $\sim100$ pc away, the quantity 
$n_{\rm e} t$ would increase by at least $2\times10^{11}$ cm$^{-3}$~s. 
This would be sufficient to change the emission line structure significantly, 
but no difference is found between the near and far sides of the GC. 
These facts are against a GC origin of the hot plasmas. 
An even more fundamental problem is that the observed velocity, either 
$\sim3300$ km~s$^{-1}$ or $\sim5000$ km~s$^{-1}$, greatly exceeds the sound 
speed in a 14-keV plasma ($\sim1400$ km s$^{-1}$). This makes it very 
unlikely that the line broadening is due to plasma motions. 

The above discussions present critical problems concerning the thermal 
plasma origin of the X-ray emission near to the GC. 
An alternative interpretation has been presented by Tanaka, Miyaji, and 
Hasinger (1999), in which X-ray line emission through charge-exchange 
interactions of low-energy cosmic-ray heavy ions was considered. When a 
low-energy cosmic-ray ion of charge $q$ undergoes a charge-exchange 
interaction with interstellar matter (mostly hydrogen), it captures an 
electron and becomes an ion of charge $q$--1 in an excited state, which 
subsequently settles to the ground state by emitting characteristic X-rays. 
Because the charge exchange has a very large cross section at low energies 
(e.g. Phaneuf et al. 1987), it is an efficient process for 
producing X-ray emission lines.
The authors point out the similarity between the observed X-ray line intensity 
ratios for various elements and the cosmic and cosmic-ray abundance ratios. 
Also, the observed line broadening is explained in terms of a sharp energy 
dependence of the charge-exchange interaction cross section. 
The cross section is very steep, $\propto E^{-4.5}$, 
when $Eq^{-1/2}\gg 25$ keV/nucleon, where $E$ is the ion energy per 
nucleon. However, it is essentially constant for 
$Eq^{-1/2}<25$ keV/nucleon. Therefore, as ions slow down by ionizing the 
ambient medium, they undergo charge-exchange interactions preferentially 
at energies of around 25$q^{1/2}$ keV/nucleon. For helium-like and 
hydrogen-like iron ions, this energy corresponds to $\sim5000$ km s$^{-1}$,  
fully consistent with that observed.  

A crucial test can be provided by a precise measurement of the widths of 
the lines from the lighter elements, since the line width is expected to be 
proportional to $q^{1/2}$ in this scenario. On the other hand, the line width 
should be the same for all the lines, if it is due to plasma motions. 
New-generation high-performance X-ray astronomy missions may provide an  
answer.

We have shown that a hot plasma of the GC origin is unlikely, yet it shows a 
temperature that is too high for a locally produced plasma by supernovae. 
Moreover, the unusual properties of the observed X-ray emission cast doubt 
on a thermal origin. Tanaka, Miyaji, and Hasinger (1999) also pointed out 
that the X-ray emission along the Galactic ridge shows a very similar 
spectrum to that near to the GC, suggesting the same origin for the 
extended X-ray emission along the entire Galactic plane. 
Identifying the origin of this X-ray emission is very important. 
It is probable that other spiral galaxies also exhibit similar X-ray emission.  
Such observations with the recent high-sensitivity missions will also 
provide valuable clues to solve the question.












\section*{References} 
\small

\noindent
Eckart A., Genzel R. 1996, Nature 383, 415

\noindent
Gendreau K. C., Mushotzky R.,  Fabian A. C., Holt S. S., Kii T., Serlemitsos 
P. J., Ogasaka Y., Tanaka Y. et al. 1995, PASJ 47, L5

\noindent
Genzel R., Hollenbach D., Townes C. H. 1994, Reports of Progress in Physics 
57, 417

\noindent
Koyama K., Maeda Y., Sonobe T., Takeshima T., Tanaka Y., Yamauchi S. 1996, 
PASJ 48, 249 (Paper I)

\noindent
Masai K., 1984, AP\&SS 98, 367

\noindent
Mezger P. G., Duschl W. J., Zylka R. 1996, A\&AR 7, 289

\noindent
Murakami H.,  Koyama K., Sakano M., Tsujimoto M., Maeda Y. 2000, ApJ 534, 283

\noindent
Phaneuf R. A., Janev R. K., Hunter H. T. 1987, in Nuclear Fusion (International 
Atomic Energy Agency, Vienna) p13 

\noindent
Tanaka Y., Inoue H., Holt S. S. 1994, PASJ 46, L37

\noindent
Tanaka Y., Miyaji T., Hasinger G. 1999, Astron. Nachr. 320, 181

\noindent
Yamauchi S., Kawada M., Koyama K., Kunieda H., Tawara Y., Hatsukade I. 1990, 
ApJ 365, 532\\

\clearpage

\end{document}